\documentclass{moriond}

\def\be{\begin{equation}}
\def\ee{\end{equation}}
\def\bea{\begin{eqnarray}}
\def\eea{\end{eqnarray}}

\usepackage{amsmath}

\newcommand{\Photo}{}

\begin{document}
\vspace*{4cm}
\title{Rare and very rare decays at the LHCb experiment}

\author{H. Tilquin on behalf of the LHCb collaboration }

\address{Department of Physics, Blackett Laboratory, Imperial College London \\
Prince Consort Road, South Kensington, London SW7 2BW, United Kingdom}
\maketitle\abstracts{
Rare and very rare decays of third-generation particles, including $b$-hadrons and $\tau$ leptons, provide sensitive probes of physics beyond the Standard Model (SM). Unlike direct searches limited by collider energies, they probe new physics at much higher energy scales. Many of these decays have SM-predicted branching fractions below the sensitivity of current detectors. These proceedings report on recent LHCb searches, including several first searches and results setting the most stringent limits to date. In particular, searches for $b \to s \tau^+\tau^-$, $b \to s \tau^\pm e^\mp$, $b \to s \mu^\pm e^\mp$, and $\tau^- \to \mu^-\mu^+\mu^-$ are presented, alongside searches for lepton-number-violating processes and loop-suppressed annihilation decays.
}

\section{Rare and very rare searches at LHCb}
Rare decays of third-generation particles, including heavy-flavoured hadrons ($b$-hadrons) and $\tau$ leptons, are excellent probes of the SM and of physics beyond it. These processes receive contributions from virtual particles in higher-order loop diagrams, making them sensitive to energy scales beyond the direct reach of collider experiments. Progress is driven mainly by increased experimental precision and larger datasets rather than higher collider energies.

In the SM, flavour-changing neutral currents (FCNC) are forbidden at tree level, while lepton-flavour violation (LFV) is forbidden for massless neutrinos and remains highly suppressed when neutrino masses are introduced, and lepton-number violation (LNV) is forbidden. Rare decays therefore provide particularly clean probes of new physics (NP), as potential NP contributions could lead to significantly enhanced decay rates relative to SM expectations.

The results presented here are based on the Run 2 dataset and, where indicated, the combined Run 1 and Run 2 datasets collected by the LHCb experiment, corresponding to integrated luminosities of 5.4 fb$^{-1}$ (collected during the data-taking years 2016–2018 of Run 2) and 9 fb$^{-1}$ (Runs 1+2), for $pp$ collisions at $\sqrt{s}=7$ TeV, 8 TeV (Run 1) and 13 TeV (Run 2).

The LHCb detector~\cite{LHCb:2008vvz} is a single-arm forward spectrometer covering the pseudorapidity range $2 < \eta < 5$, optimised for the study of heavy-flavour hadrons. Its excellent vertex resolution and particle identification capabilities, particularly for muons~\cite{Archilli:2013npa}, enable efficient background suppression. This is essential for probing decays with very low branching fractions and for (partially) reconstructing final states with missing energy, such as $B$-meson decays to $\tau$ leptons.

These proceedings focus on searches for rare and very rare decays, including the SM-suppressed FCNC processes $b \to s \tau^+\tau^-$ which remain experimentally challenging due to the presence of $\tau$ leptons. They also cover LFV $b \to s \tau^\pm e^\mp$, $b \to s \mu^\pm e^\mp$ and $\tau^- \to \mu^-\mu^+\mu^-$ transitions, as well as LNV decays and loop-suppressed annihilation processes. Many of these processes are predicted to have branching fractions well below current experimental sensitivity. 
Observation at current sensitivities of LFV or LNV, or enhancements in $b\to s\tau^+\tau^-$ processes, would thus indicate physics beyond the SM, complementing direct searches for new particles.

\section{Analysis techniques}

Searches for rare and forbidden decays at the LHCb experiment are typically constrained by background contributions due to the extremely low branching fractions of the processes under study. The following provides a general overview of the strategies employed to reduce backgrounds; individual analyses often include additional channel-specific optimisations.

Three main types of background are considered:

\begin{itemize}
    \item Combinatorial background: random combinations of tracks that do not originate from the same decay but mimic the signal topology.
    \item Partially reconstructed background: decays in which one or more final-state particles are not reconstructed.
    \item Misidentified (misID) background: particles incorrectly identified as another species, for example a pion reconstructed as a muon.
\end{itemize}

Backgrounds are suppressed using a combination of selection requirements, including particle identification and multivariate classifiers such as boosted decision trees (BDTs)~\cite{Breiman,AdaBoost}.  
Signal yields are extracted using fits to either the reconstructed invariant mass of the candidate or the output of the multivariate classifier, depending on the channel. The templates used in these fits can be analytic or non-parametric, as appropriate to the signal and background distributions.  
In cases where no statistically significant signal is observed, upper limits on branching fractions are set using the CL$_\mathrm{s}$ method~\cite{CLs}, providing constraints on possible contributions from NP.

\section{Searches for \texorpdfstring{$\boldsymbol{b\to s \tau^+\tau^-}$}{b to s tau tau} transitions}

In the SM, $b \to s \tau^+\tau^-$ processes are highly suppressed with branching fractions of $\mathcal{O}(10^{-7})$, but several NP scenarios predict sizeable enhancements~\cite{Capdevila:2017iqn} linked to the discrepancies observed in $R(D^{(*)})$~\cite{HFLAV:2024ctg}. The present searches~\cite{LHCb:2025lcw} focus on $B^0 \to K^+ \pi^- \tau^+\tau^-$ and $B_s^0 \to K^+ K^- \tau^+\tau^-$, with $\tau^+ \to \mu^+ \bar{\nu}_\tau \nu_\mu$.  
These channels are challenging due to the presence of multiple neutrinos, which prevents full invariant-mass reconstruction. The $B$-meson decay vertex is reconstructed from the two charged hadrons, and muons are selected using LHCb's excellent muon identification.

Backgrounds include $B\to D^{(*)}\bar{D}^{(*)}X$ ($X=\phi(1020), K^*(892)^0$) decays in which partially reconstructed $D$-meson decays mimic the signal, misID candidates,  combinatorial contributions and semileptonic decays. Suppression is achieved through selection criteria on the reconstructed squared missing mass and squared ditau mass, particle identification, as well as a multiclass boosted decision tree (BDT) trained to separate signal from combinatorial and semileptonic backgrounds in bins of dihadron mass. Residual misID background is included in the final fits.

Candidates with a signal BDT score exceeding either background score are fitted, with templates derived from simulation and control data using kernel density estimations~\cite{Cranmer:2000du}. No statistically significant signal is observed, and upper limits are set. Limits are set in bins of dihadron mass as well as on $\mathcal{B}(B_s^0 \to K^-\pi^+ \tau^+\tau^-)$ (see Table~\ref{tab:Table1}), and on intermediate resonances~\cite{LHCb:2025lcw}:
\begin{equation}
    \begin{split}
        \mathcal{B}(B^0 \to K^*(892)^0\tau^+\tau^-) &< 2.8 \, (2.5)\times 10^{-4} \textrm{ at $95\%$ ($90\%$) CL}, \\
        \mathcal{B}(B_s^0 \to \phi(1020)\tau^+\tau^-) &< 4.7 \,(4.1) \times 10^{-4} \textrm{ at $95\%$ ($90\%$) CL}.
    \end{split}
\end{equation}

\begin{table}[htb]
  \caption{Upper limits on $b\to s\tau^+\tau^-$ and $b\to d\tau^+\tau^-$ branching fractions in bins of dihadron mass.}
\begin{center}
\begin{tabular}{|c c c c c c|}
    \hline
    CL  & \multicolumn{4}{c}{Upper limit on $\mathcal{B}(B^0\to K^+\pi^-\tau^+\tau^-)$} & \\
    \hline 
    $m_{K^+\pi^-}$ (MeV) & $[792, 992]$ & $[992, 1330]$ & $[1330, 1530]$ & $[1530, 1726]$ & \\
    \hline 
   $90\%$ & $1.4\times10^{-4}$ & $2.7\times10^{-5}$ & $1.0\times10^{-5}$ & $ 2.7\times10^{-6}$  &  \\
   $95\%$ & $1.6\times10^{-4}$ & $3.4\times10^{-5}$ & $1.1\times10^{-5}$ & $ 3.3\times10^{-6}$ & \\
    \hline
    CL  & \multicolumn{4}{c}{Upper limit on $\mathcal{B}(B_s^0\to K^+ K^-\tau^+\tau^-)$} &  \\
    \hline 
   $m_{K^+ K^-}$ (MeV) &  $[980, 1060]$ & $[1060, 1200]$ & $[1200, 1400]$ & $[1400, 1600]$ & $[1600, 1813]$  \\
     \hline 
   $90\%$ & $2.0\times10^{-4}$ & $1.3\times10^{-4}$ & $1.2\times10^{-4}$ & $6.8\times10^{-5}$ & $3.2\times10^{-5}$ \\
   $95\%$ & $2.3\times10^{-4}$ & $1.5\times10^{-4}$ & $1.4\times10^{-4}$ & $7.6\times10^{-5}$ & $3.6\times10^{-5}$ \\
    \hline
    CL  & \multicolumn{4}{c}{Upper limit on $\mathcal{B}(B_s^0\to K^-\pi^+\tau^+\tau^-$)} & \\
    \hline 
    $m_{K^+\pi^-}$ (MeV) & $[792, 992]$ & $[992, 1330]$ & $[1330, 1530]$ & $[1530, 1726]$ & \\
    \hline 
   $90\%$ & $6.5\times10^{-4}$ & $1.2\times10^{-4}$ & $5.1\times10^{-5}$ & $1.7\times10^{-5}$ &  \\
   $95\%$ & $7.3\times10^{-4}$ & $1.5\times10^{-4}$ & $6.2\times10^{-5}$ & $2.1\times10^{-5}$ &  \\
    \hline
  \end{tabular}\end{center}
\label{tab:Table1}
\end{table}

In addition, the results are recast as constraints on Wilson coefficients using the model and assumptions of Ref.~\cite{Capdevila:2017iqn} (see Table~\ref{tab:Table2}). In this scenario, the branching fractions of $b\to s \tau^+\tau^-$ transitions are directly proportional to the square of the shift $\Delta$ in the Wilson coefficients $\mathcal{C}_{9(10)}^{\tau\tau}$, where $\mathcal{C}^{\tau\tau}_{9(10)} = \mathcal{C}^{\textrm{SM}}_{9(10)} \mp \Delta$.

\begin{table}[!htb]
  \caption{
  Upper limit on $\Delta^2$ at $90\%$ and $95\%$ CL, assuming $\mathcal{C}^{\textrm{NP}}_{9(10)} = \mathcal{C}^{\textrm{SM}}_{9(10)} \mp \Delta$ (see Ref.~\protect\cite{Capdevila:2017iqn}).}
\begin{center}
\begin{tabular}{|c c c |}
    \hline
    CL & $B^0\to K^+\pi^-\tau^+\tau^-$ & $B_s^0\to K^+K^-\tau^+\tau^-$ \\
    \hline 
   $90\%$ & $2.5\times10^4$ & $4.5\times10^4$  \\
   $95\%$ & $2.9\times10^4$ & $5.2\times10^4$ \\
    \hline
  \end{tabular}\end{center}
  \label{tab:Table2}
\end{table}

\begin{figure}[!htb]
    \centering
    \includegraphics[width=0.5\linewidth]{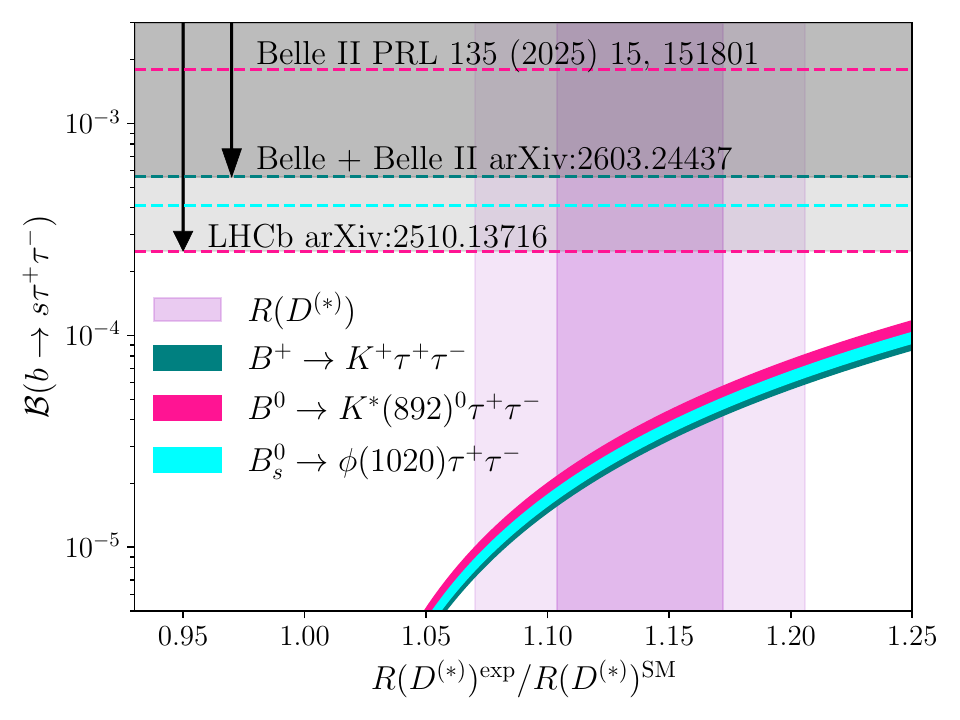}
    \caption{
    NP predictions from Ref.\protect\cite{Capdevila:2017iqn}, using the latest $R(D^{(*)})$ results from HFLAV~\protect\cite{HFLAV:2024ctg}, with the latest Belle and Belle II~\protect\cite{Belle-II:2025lwo}~\protect\cite{Belle-II:2026ism} and LHCb~\protect\cite{LHCb:2025lcw} limits (at $90\%$ CL) overlaid.}
    \label{fig:b2stautau_status}
\end{figure}

These results represent either the first searches or the world’s most stringent constraints on $b \to s \tau^+\tau^-$ transitions to date (see Fig.~\ref{fig:b2stautau_status}). The dominant systematic uncertainty arises from the limited size of the data-derived background templates, which suggests improvements with larger datasets collected in current and upcoming data-taking periods.

\section{Lepton Flavour Violation Searches}

LFV in charged lepton processes is essentially forbidden in the SM. Many NP models, however, predict significant enhancements, with branching fractions potentially reaching $\mathcal{O}(10^{-10})$~\cite{Duraisamy:2016gsd}. Any observation of LFV would therefore constitute an unambiguous sign of physics beyond the SM.

\subsection{Searches for \texorpdfstring{$B^0\to K^*(892)^0\tau^\pm e^\mp$}{B0 to K* tau e}}
\begin{figure}[!ht]
    \centering
    \includegraphics[width=0.82\linewidth]{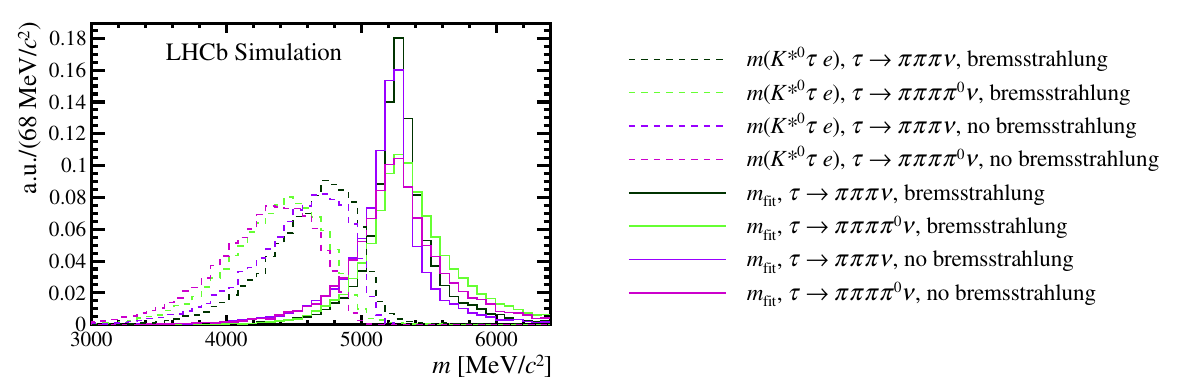}
    \caption{Reconstructed mass $m$ and constrained reconstructed mass $m_{\rm fit}$ for final states with and without bremsstrahlung effects, and with or without an additional $\pi^0$ stemming from the $\tau$ decay~\protect\cite{LHCb:2025eyf}.}
    \label{fig:mfit}
\end{figure}

The LHCb collaboration has recently probed LFV in $b \to s \tau^\pm \ell^\mp$ transitions with the decay $B^0 \to K^{*}(892)^0 \tau^\pm e^\mp$, using three-prong pionic $\tau$ decays~\cite{LHCb:2025eyf}. Experimentally, it is challenging due to the presence of a $\tau$ lepton, which decays in this case to charged tracks and a neutrino. With the $B^0$ decay vertex reconstructed using the $K^*(892)^0$ and electron, and the $\tau$ candidate constrained using the known $\tau$ mass and the reconstructed $\tau$ decay vertex from three-prong decays, a constrained reconstructed mass $m_{\rm fit}$ is obtained (see Fig.~\ref{fig:mfit} for a comparison with the unconstrained mass). Bremsstrahlung effects are taken into account in the modelling of the signal shape.

Backgrounds include combinatorial tracks, non-isolated candidates, misidentified hadrons, and charm background from decays of $D$-meson, which have a lifetime similar to that of $\tau$ leptons. Background suppression is achieved through sequential selection cuts and multivariate classifiers, with the dominant systematic uncertainty arising from the choice of the background control region. The signal yield is extracted by fitting the constrained mass $m_{\rm fit}$. No significant signal is observed. Upper limits, determined separately for each charge combination, are set~\cite{LHCb:2025eyf}:
\begin{equation}
    \begin{split}
        \mathcal{B}(B^0 \to K^{*0}\tau^- e^+) & < 5.9 \, (7.1) \times 10^{-6} \textrm{ at 90\% (95\%) CL}, \\
        \mathcal{B}(B^0 \to K^{*0}\tau^+ e^-) & < 4.9 \, (5.9) \times 10^{-6} \textrm{ at 90\% (95\%) CL}.
    \end{split}
\end{equation}

These results set the most stringent constraints to date on $b \to s \tau^\pm e^\mp$ transitions, improving on previous searches~\cite{Belle:2022pcr,Belle:2025sgv}.

\subsection{Searches for \texorpdfstring{$B^+\to \pi^+\mu^\pm e^\mp$}{B+ to pi mu e}}

The decay $B^+ \to \pi^+ \mu e$ provides a complementary probe of LFV with a fully reconstructible final state. The analysis~\cite{LHCb-PAPER-2026-013} employs two BDTs: the first reduces combinatorial background using the upper mass sideband, while the second suppresses remaining combinatorial and partially reconstructed backgrounds using the lower mass sideband. The dominant systematic uncertainty arises from the modelling of the background components. The signal yield is extracted from a fit to the reconstructed invariant mass. No signal is observed, and an upper limit is set~\cite{LHCb-PAPER-2026-013}:
\begin{equation}
 \mathcal{B}(B^+\to \pi^+\mu^\pm e^\mp) < 1.8 \, (2.2)\times10^{-9} \textrm{ at 90\% (95\%) CL},
\end{equation}
using the full LHCb Run~1 and Run~2 dataset. This result improves on previous bounds~\cite{BaBar:2007xeb} by two orders of magnitude.

\subsection{Searches for \texorpdfstring{$\tau^- \to \mu^- \mu^+ \mu^-$}{tau to mu mu mu}}

In the SM (including neutrino masses), the branching fraction of the $\tau^- \to \mu^- \mu^+ \mu^-$ decay is negligible~\cite{Blackstone:2019njl} ($\sim 10^{-55}$) but NP models predict enhancements up to $10^{-10}$--$10^{-8}$~\cite{Cvetic:2002jy,Yue:2002ja}.
Backgrounds arise from combinatorial tracks and hadrons misidentified as muons, including decays of $D_{(s)}^-$ mesons. Two BDTs are used~\cite{LHCb:2026eod}: the first suppresses combinatorial background, while the second targets misID backgrounds. Signal yields are extracted from a fit to the three-muon invariant mass. 
No significant signal is observed. The resulting upper limit is~\cite{LHCb:2026eod}:
\begin{equation}
\mathcal{B}(\tau^-\to\mu^-\mu^+\mu^-) < 1.9 \, (2.3) \times 10^{-8} \textrm{ at 90\% (95\%) CL},
\end{equation}
and is on par with the limit recently obtained by the Belle II collaboration~\cite{Belle-II:2024sce}. The dominant systematic uncertainty arises from external inputs, such as the knowledge of $\tau$ production fractions from $D_s$ decays.

\section{Lepton Number Violation Searches}
\begin{figure}[!htb]
    \centering
    \includegraphics[width=0.31\linewidth]{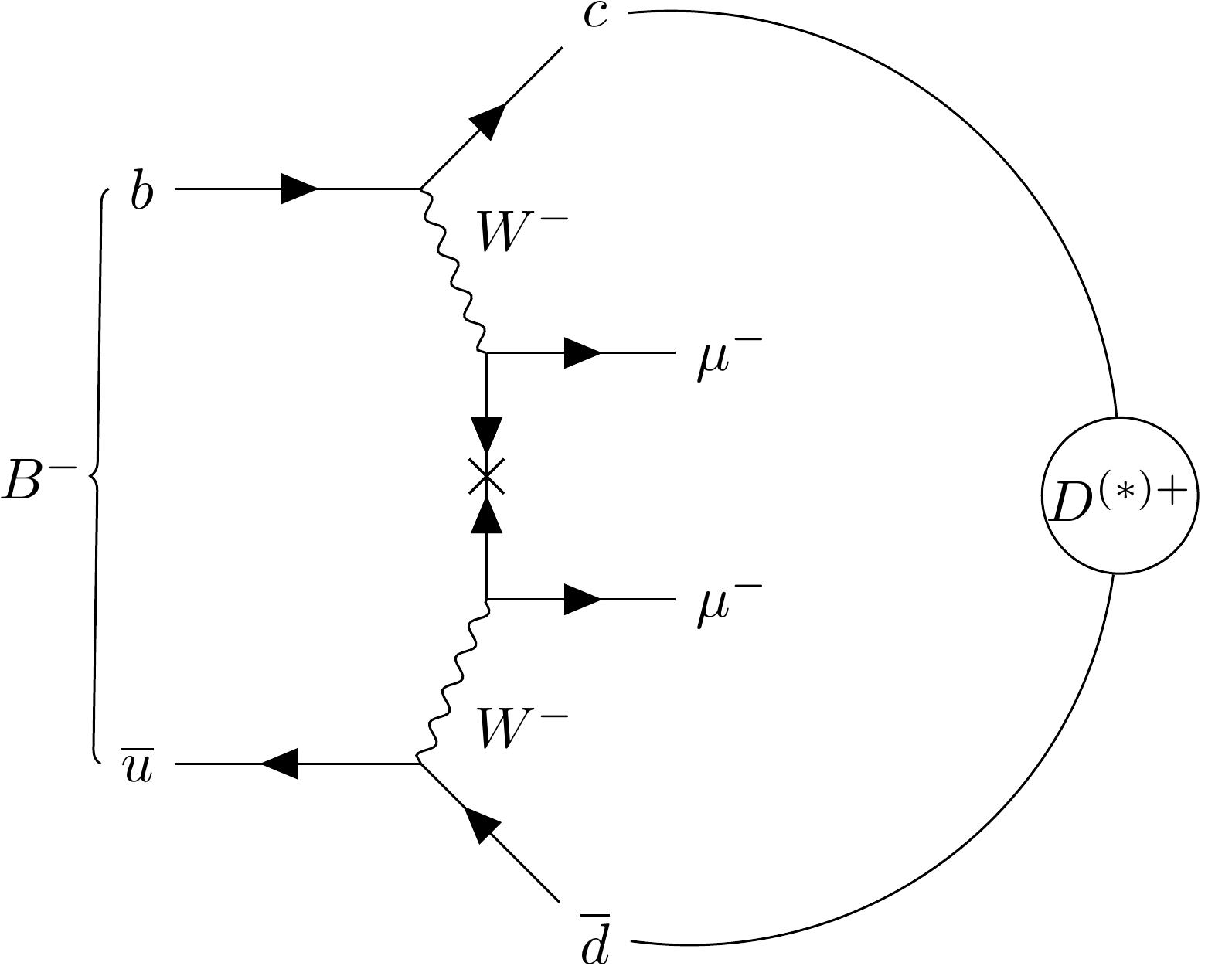}
    \caption{Feynman diagram for $B^-\to D^{(*)+}\mu^-\mu^-$ decays occuring with a Majorana neutrino~\protect\cite{LHCb:2026evs}.}
    \label{fig:Fig1_b2dmumu}
\end{figure}

Decays such as $B^-\to D^{(*)+}\mu^-\mu^-$ (see Fig.~\ref{fig:Fig1_b2dmumu}) would violate lepton number conservation. In the presence of Majorana neutrinos (which are hypothetical particles identical to their anti-particles), the branching fractions of such decays would be $\mathcal{O}(10^{-23})$--$\mathcal{O}(10^{-22})$~\cite{Cvetic:2010rw}, far below experimental sensitivity. Nevertheless, such searches provide useful constraints on LNV processes.

Searches for $B^- \to D^{(*)+} \mu^- \mu^-$ decays have been conducted at the LHCb experiment~\cite{LHCb:2026evs}. A BDT is used to separate simulated signal from combinatorial background. Additional backgrounds arise from pion misidentification to muons and pions decaying in-flight to muons ($\pi^+\to\mu^+\nu_\mu$), which form a non-negligible contribution and are included in the final fit. The dominant systematic uncertainty arises from the decay model of the signal, which is assumed to be generated uniformly in phase space. The signal yield is extracted from a fit to the reconstructed invariant mass. No significant signal is observed, and upper limits are set~\cite{LHCb:2026evs}:
\begin{equation}
\begin{split}
    \mathcal{B}(B^-\to D^+\mu^-\mu^-) &< 3.8 \, (4.6) \times 10^{-8} \textrm{ at 90\% (95\%) CL}, \\
    \mathcal{B}(B^- \to D^{*+}\mu^-\mu^-) &< 4.5 \, (5.9) \times 10^{-8} \textrm{ at 90\% (95\%) CL}.
\end{split}
\end{equation}

These results improve on previous searches~\cite{LHCb:2012pcm} by an order of magnitude.

\section{Loop-suppressed Annihilation Search}
The decay $B^0 \to \phi \phi$ is both OZI- and Cabibbo-suppressed in the SM. Its predicted branching fraction lies between $0.5\times10^{-8}$ and $5\times10^{-8}$~\cite{Bar-Shalom:2002icz,Li:2013uaa}.

Backgrounds arise primarily from the low-mass tail of $B_s^0 \to \phi \phi$ decays, where kaons interact hadronically with the tracking system or decay in flight to muons. Additional combinatorial background originates from $D_s^-$ decays and generic processes. Backgrounds are suppressed using multivariate classifiers exploiting kinematic variables and by applying requirements on the $\phi$ candidate masses and on the absence of associated activity in the muon stations~\cite{LHCb:2025gog}.

No significant signal is observed in a fit to the reconstructed invariant mass. Using the full Run~1 and Run~2 dataset, an upper limit is set~\cite{LHCb:2025gog}, with the dominant systematic uncertainty arising from the limited knowledge of the branching fraction of $B_s^0 \to \phi \phi$:
\begin{equation}
 \mathcal{B}(B^0 \to \phi\phi) < 1.3 \, (1.4) \times 10^{-8} \textrm{ at 90\% (95\%) CL}.
\end{equation}

 This result improves on the previous limit~\cite{LHCb:2019jgw} by a factor of two.

\section{Conclusion}
A comprehensive programme of searches for rare and very rare decays has been carried out at the LHCb experiment using Run~1 and Run~2 data, yielding the most stringent limits to date for several decay modes. Looking ahead, Run~3 and beyond will provide a much larger dataset which, combined with the upgraded detector~\cite{LHCb:2023hlw} with a fully software-based trigger, will improve sensitivity to rare and forbidden decays and enable more stringent tests of the SM.

\section*{References}
\bibliography{moriond}

@article{LHCb:2025eyf,
    author = "Aaij, R. and others",
    collaboration = "LHCb",
    eprint = "2506.15347",
    archivePrefix = "arXiv",
    primaryClass = "hep-ex",
    reportNumber = "LHCb-PAPER-2025-005, CERN-EP-2025-107",
    doi = "10.1007/JHEP11(2025)172",
    journal = "JHEP",
    volume = "11",
    pages = "172",
    year = "2025",
    note = "\href{https://arxiv.org/abs/2506.15347}{arXiv:2506.15347}, \href{https://doi.org/10.1007/JHEP11(2025)172}{doi:10.1007/JHEP11(2025)172}"
}

@article{LHCb-PAPER-2026-013,
    author = "Aaij, R. and others",
    collaboration = "LHCb",
    reportNumber = "LHCb-PAPER-2026-013, CERN-EP-2026-093",
    eprint = "2604.08396",
    archivePrefix = "arXiv",
    primaryClass = "hep-ex",
    doi = "",
    journal = "",
    volume = "",
    pages = "",
    year = "2026",
    note = "\href{https://arxiv.org/abs/2604.08396}{arXiv:2604.08396}"
}

@article{LHCb:2025lcw,
    author = "Aaij, R. and others",
    collaboration = "LHCb",
    eprint = "2510.13716",
    archivePrefix = "arXiv",
    primaryClass = "hep-ex",
    reportNumber = "LHCb-PAPER-2025-048, CERN-EP-2025-224",
    year = "2025",
    note = {\href{https://arxiv.org/abs/2510.13716}{arXiv:2510.13716}}
}

@article{LHCb:2026eod,
    author = "Aaij, R. and others",
    collaboration = "LHCb",
    eprint = "2601.20785",
    archivePrefix = "arXiv",
    primaryClass = "hep-ex",
    reportNumber = "LHCb-PAPER-2025-052, CERN-EP-2026-003",
    year = "2026",
    note = "\href{https://arxiv.org/abs/2601.20785}{arXiv:2601.20785}"
}

@article{LHCb:2026evs,
    author = "Aaij, R. and others",
    collaboration = "LHCb",
    eprint = "2601.07657",
    archivePrefix = "arXiv",
    primaryClass = "hep-ex",
    reportNumber = "LHCb-PAPER-2025-033, CERN-EP-2025-267",
    year = "2026",
    note = "\href{https://arxiv.org/abs/2601.07657}{arXiv:2601.07657}"

}

@article{HFLAV:2024ctg,
    author = "Banerjee, S. and others",
    journal = {Phys. Rev.},
    volume = {D113},
    issue = {1},
    pages = {012008},
    doi = {10.1103/x87q-tld5},
    eprint = "2411.18639",
    archivePrefix = "arXiv",
    primaryClass = "hep-ex",
    year = "2026",
    note = "\href{https://arxiv.org/abs/2411.18639}{arXiv:2411.18639}, \href{https://doi.org/10.1103/x87q-tld5}{doi:10.1103/x87q-tld5}"

}

@article{Capdevila:2017iqn,
    author = "Capdevila, B. and Crivellin, A. and Descotes-Genon, S. and Hofer, L. and Matias, J.",
    eprint = "1712.01919",
    archivePrefix = "arXiv",
    primaryClass = "hep-ph",
    reportNumber = "PSI-PR-17-19, LPT-ORSAY-17-74",
    doi = "10.1103/PhysRevLett.120.181802",
    journal = "Phys. Rev. Lett.",
    volume = "120",
    number = "18",
    pages = "181802",
    year = "2018",
    note = "\href{https://arxiv.org/abs/1712.01919}{arXiv:1712.01919}, \href{https://doi.org/10.1103/PhysRevLett.120.181802}{doi:10.1103/PhysRevLett.120.181802}"
}

@article{Duraisamy:2016gsd,
    author = "Duraisamy, M. and Sahoo, S. and Mohanta, R.",
    eprint = "1610.00902",
    archivePrefix = "arXiv",
    primaryClass = "hep-ph",
    doi = "10.1103/PhysRevD.95.035022",
    journal = "Phys. Rev. D",
    volume = "95",
    number = "3",
    pages = "035022",
    year = "2017",
    note = "\href{https://arxiv.org/abs/1610.00902}{arXiv:1610.00902}, \href{https://doi.org/10.1103/PhysRevD.95.035022}{doi:10.1103/PhysRevD.95.035022}"

}

@article{Blackstone:2019njl,
    author = "Blackstone, P. and Fael, M. and Passemar, E.",
    eprint = "1912.09862",
    archivePrefix = "arXiv",
    primaryClass = "hep-ph",
    reportNumber = "P3H-19-54, TPP19-48, SI-HEP-2019-22, JLAB-THY-19-3123",
    doi = "10.1140/epjc/s10052-020-8059-7",
    journal = "Eur. Phys. J. C",
    volume = "80",
    number = "6",
    pages = "506",
    year = "2020",
     note = "\href{https://arxiv.org/abs/1912.09862}{arXiv:1912.09862}, \href{https://doi.org/10.1140/epjc/s10052-020-8059-7}{doi:10.1140/epjc/s10052-020-8059-7}"
}

@article{Cvetic:2002jy,
    author = "Cvetic, G. and Dib, C. and Kim, C. S. and Kim, J. D.",
    eprint = "hep-ph/0202212",
    archivePrefix = "arXiv",
    reportNumber = "USM-TH-123",
    doi = "10.1103/PhysRevD.66.034008",
    journal = "Phys. Rev. D",
    volume = "66",
    pages = "034008",
    year = "2002",
    note = "[Erratum: Phys.Rev.D 68, 059901 (2003)] \href{https://arxiv.org/abs/hep-ph/0202212}{arXiv:hep-ph/0202212}, \href{https://doi.org/10.1103/PhysRevD.66.034008}{doi:110.1103/PhysRevD.66.034008}"

}

@article{Yue:2002ja,
    author = "Yue, C. and Zhang, Y. and Liu, L.",
    eprint = "hep-ph/0209291",
    archivePrefix = "arXiv",
    doi = "10.1016/S0370-2693(02)02781-8",
    journal = "Phys. Lett. B",
    volume = "547",
    pages = "252--256",
    year = "2002",
     note = "\href{https://arxiv.org/abs/hep-ph/0209291}{arXiv:hep-ph/0209291}, \href{https://doi.org/10.1016/S0370-2693(02)02781-8}{doi:10.1016/S0370-2693(02)02781-8}"

}

@article{CLs,
  author="A. L. Read",
  journal="J. Phys.",
  volume="G28",
  pages="2693",
  doi="10.1088/0954-3899/28/10/313",
  year="2002",
     note = "\href{https://doi.org/10.1088/0954-3899/28/10/313}{doi:110.1088/0954-3899/28/10/313}"

}

@book{Breiman,
  author = 	 {Breiman, L. and Friedman, J. H. and Olshen,
                  R. A. and Stone, C. J.},
  title = 	 {Classification and regression trees},
  publisher = 	 {Wadsworth international group},
  year = 	 {1984},
  address = 	 {Belmont, California, USA},
}

@article{AdaBoost,
    author = "Freund, Y. and Schapire, R. E.",
  journal = 	 "J. Comput. Syst. Sci.",
  volume         = "55",
  pages          = "119",
  year = 	 {1997},
  doi            = "10.1006/jcss.1997.1504",
     note = "\href{https://doi.org/10.1006/jcss.1997.1504}{doi:10.1006/jcss.1997.1504}"

}

@article{Cranmer:2000du,
    author = "Cranmer, K. S.",
    eprint = "hep-ex/0011057",
    archivePrefix = "arXiv",
    doi = "10.1016/S0010-4655(00)00243-5",
    journal = "Comput. Phys. Commun.",
    volume = "136",
    pages = "198--207",
    year = "2001",
     note = "\href{https://doi.org/10.1016/S0010-4655(00)00243-5}{doi:10.1016/S0010-4655(00)00243-5}"

}

@article{Belle-II:2026ism,
    author = "Abumusabh, M. and others",
    collaboration = "Belle-II, Belle",
    eprint = "2603.24437",
    archivePrefix = "arXiv",
    primaryClass = "hep-ex",
    reportNumber = "Belle II preprint 2026-003, KEK preprint 2025-43",
    year = "2026",
     note = "\href{https://arxiv.org/abs/2603.24437}{arXiv:2603.24437}"

}

@article{Bar-Shalom:2002icz,
    author = "Bar-Shalom, S. and Eilam, G. and Yang, Y.",
    eprint = "hep-ph/0201244",
    archivePrefix = "arXiv",
    doi = "10.1103/PhysRevD.67.014007",
    journal = "Phys. Rev. D",
    volume = "67",
    pages = "014007",
    year = "2003",
     note = "\href{https://arxiv.org/abs/hep-ph/0201244}{arXiv:hep-ph/0201244}, \href{https://doi.org/10.1103/PhysRevD.67.014007}{doi:10.1103/PhysRevD.67.014007}"

}

@article{Li:2013uaa,
    author = "Li, Y.",
    eprint = "1311.2664",
    archivePrefix = "arXiv",
    primaryClass = "hep-ph",
    doi = "10.1103/PhysRevD.89.014003",
    journal = "Phys. Rev. D",
    volume = "89",
    number = "1",
    pages = "014003",
    year = "2014",
     note = "\href{https://arxiv.org/abs/1311.2664}{arXiv:1311.2664}, \href{https://doi.org/10.1103/PhysRevD.89.014003}{doi:10.1103/PhysRevD.89.014003}"

}

@article{Cvetic:2010rw,
    author = "Cvetic, G. and Dib, Claudio and Kang, Sin Kyu and Kim, C. S.",
    eprint = "1005.4282",
    archivePrefix = "arXiv",
    primaryClass = "hep-ph",
    doi = "10.1103/PhysRevD.82.053010",
    journal = "Phys. Rev. D",
    volume = "82",
    pages = "053010",
    year = "2010",
     note = "\href{https://arxiv.org/abs/1005.4282}{arXiv:1005.4282}, \href{https://doi.org/10.1103/PhysRevD.82.053010}{doi:10.1103/PhysRevD.82.053010}"
}

@article{Belle-II:2024sce,
    author = "Adachi, I. and others",
    collaboration = "Belle-II",
    eprint = "2405.07386",
    archivePrefix = "arXiv",
    primaryClass = "hep-ex",
    reportNumber = "Belle II Preprint 2024-012 KEK Preprint 2024-6",
    doi = "10.1007/JHEP09(2024)062",
    journal = "JHEP",
    volume = "09",
    pages = "062",
    year = "2024",
     note = "\href{https://arxiv.org/abs/2405.07386}{arXiv:2405.07386}, \href{https://doi.org/10.1007/JHEP09(2024)062}{doi:10.1007/JHEP09(2024)062}"
}

@article{LHCb:2008vvz,
    author = "Alves, Jr., A. Augusto and others",
    collaboration = "LHCb",
    reportNumber = "LHCb-DP-2008-001",
    doi = "10.1088/1748-0221/3/08/S08005",
    journal = "JINST",
    volume = "3",
    pages = "S08005",
    year = "2008",
     note = "\href{https://doi.org/10.1088/1748-0221/3/08/S08005}{doi:10.1088/1748-0221/3/08/S08005}"
}

@article{Belle-II:2025lwo,
    author = "Adachi, I. and others",
    collaboration = "Belle-II",
    eprint = "2504.10042",
    archivePrefix = "arXiv",
    primaryClass = "hep-ex",
    reportNumber = "Belle II Preprint 2025-010; KEK Preprint 2025-8",
    doi = "10.1103/v1q3-9dy8",
    journal = "Phys. Rev. Lett.",
    volume = "135",
    number = "15",
    pages = "151801",
    year = "2025",
     note = "\href{https://arxiv.org/abs/2504.10042}{arXiv:2504.10042}, \href{https://doi.org/10.1103/v1q3-9dy8}{doi:10.1103/v1q3-9dy8}"

}

@article{LHCb:2023hlw,
    author = "Aaij, R. and others",
    collaboration = "LHCb",
    eprint = "2305.10515",
    archivePrefix = "arXiv",
    primaryClass = "hep-ex",
    reportNumber = "LHCb-DP-2022-002",
    doi = "10.1088/1748-0221/19/05/P05065",
    journal = "JINST",
    volume = "19",
    number = "05",
    pages = "P05065",
    year = "2024",
     note = "\href{https://arxiv.org/abs/2305.10515}{arXiv:2305.10515}, \href{https://doi.org/10.1088/1748-0221/19/05/P05065}{doi:10.1088/1748-0221/19/05/P05065}"

}

@article{Archilli:2013npa,
    author = "Archilli, F. and others",
    eprint = "1306.0249",
    archivePrefix = "arXiv",
    primaryClass = "physics.ins-det",
    reportNumber = "CERN-LHCB-DP-2013-001, LHCB-DP-2013-001",
    doi = "10.1088/1748-0221/8/10/P10020",
    journal = "JINST",
    volume = "8",
    pages = "P10020",
    year = "2013",
     note = "\href{https://arxiv.org/abs/1306.0249}{arXiv:1306.0249}, \href{https://doi.org/10.1088/1748-0221/8/10/P10020}{doi:10.1088/1748-0221/8/10/P10020}"
}

@article{LHCb:2025gog,
    author = "Aaij, R. and others",
    collaboration = "LHCb",
    eprint = "2507.20945",
    archivePrefix = "arXiv",
    primaryClass = "hep-ex",
    reportNumber = "LHCb-PAPER-2025-018, CERN-EP-2025-150",
    doi = "10.1007/JHEP12(2025)026",
    journal = "JHEP",
    volume = "12",
    pages = "026",
    year = "2025",
     note = "\href{https://arxiv.org/abs/2507.20945}{arXiv:2507.20945}, \href{https://doi.org/10.1007/JHEP12(2025)026}{doi:10.1007/JHEP12(2025)026}"

}

@article{LHCb:2012pcm,
    author = "Aaij, R. and others",
    collaboration = "LHCb",
    eprint = "1201.5600",
    archivePrefix = "arXiv",
    primaryClass = "hep-ex",
    reportNumber = "CERN-PH-EP-2012-006, LHCB-PAPER-2011-038",
    doi = "10.1103/PhysRevD.85.112004",
    journal = "Phys. Rev. D",
    volume = "85",
    pages = "112004",
    year = "2012",
     note = "\href{https://arxiv.org/abs/1201.5600}{arXiv:1201.5600}, \href{https://doi.org/10.1103/PhysRevD.85.112004}{doi:10.1103/PhysRevD.85.112004}"
}

@article{LHCb:2019jgw,
    author = "Aaij, R. and others",
    collaboration = "LHCb",
    eprint = "1907.10003",
    archivePrefix = "arXiv",
    primaryClass = "hep-ex",
    reportNumber = "LHCb-PAPER-2019-019, CERN-EP-2019-121",
    doi = "10.1007/JHEP12(2019)155",
    journal = "JHEP",
    volume = "12",
    pages = "155",
    year = "2019",
     note = "\href{https://arxiv.org/abs/1907.10003}{arXiv:1907.10003}, \href{https://doi.org/10.1007/JHEP12(2019)155}{doi:10.1007/JHEP12(2019)155}"
}

@article{BaBar:2007xeb,
    author = "Aubert, B. and others",
    collaboration = "BaBar",
    eprint = "hep-ex/0703018",
    archivePrefix = "arXiv",
    reportNumber = "SLAC-PUB-12389, BABAR-PUB-07-002",
    doi = "10.1103/PhysRevLett.99.051801",
    journal = "Phys. Rev. Lett.",
    volume = "99",
    pages = "051801",
    year = "2007",
     note = "\href{https://arxiv.org/abs/hep-ex/0703018}{arXiv:hep-ex/0703018}, \href{https://doi.org/10.1103/PhysRevLett.99.051801}{doi:10.1103/PhysRevLett.99.051801}"

}

@article{Belle:2022pcr,
    author = "Watanuki, S. and others",
    collaboration = "Belle",
    eprint = "2212.04128",
    archivePrefix = "arXiv",
    primaryClass = "hep-ex",
    doi = "10.1103/PhysRevLett.130.261802",
    journal = "Phys. Rev. Lett.",
    volume = "130",
    number = "26",
    pages = "261802",
    year = "2023",
     note = "\href{https://arxiv.org/abs/2212.04128}{arXiv:2212.04128}, \href{https://doi.org/10.1103/PhysRevLett.130.261802}{doi:10.1103/PhysRevLett.130.261802}"
}

@article{Belle:2025sgv,
    author = "Adachi, I. and others",
    collaboration = "Belle, Belle-II",
    eprint = "2505.08418",
    archivePrefix = "arXiv",
    primaryClass = "hep-ex",
    reportNumber = "Belle II preprint: 2025-014, KEK preprint: 2025-13",
    doi = "10.1007/JHEP08(2025)184",
    journal = "JHEP",
    volume = "08",
    pages = "184",
    year = "2025",
     note = "\href{https://arxiv.org/abs/2505.08418}{arXiv:2505.08418}, \href{https://doi.org/10.1007/JHEP08(2025)184}{doi:10.1007/JHEP08(2025)184}"

}

\end{document}